\documentclass[prb,twocolumn,showpacs,preprintnumbers,amsmath,amssymb,showkeys]{revtex4}
\usepackage{amscd,graphicx,psfrag}

\renewcommand{\exp}[1]{{\rm e}^{#1}}

\newcommand{\hvect}[1]{\hat{#1}}
\newcommand{\vect}[1]{\vec{#1}}
\newcommand{\vectsym}[1]{\vec{#1}}
\renewcommand{\tensor}[1]{\boldsymbol{#1}}

\newcommand{\Z}{\mathcal{Z}}

\newcommand{\kB}{k_\text{B}}

\newcommand{\system}[1]{\mathbf{#1}}
\newcommand{\set}[1]{\{\system{#1}\}}
\newcommand{\setplusone}[2]{\left\{\system{#1}+\system{1}_{#2}\right\}}
\newcommand{\mysetminus}{\left\{\system{N}-\system{n}\right\}}
\newcommand{\order}[1]{{(\system{#1})}}
\newcommand{\orderplusone}[2]{{(\system{#1}+\system{1}_{#2})}}
\newcommand{\avg}[1]{\left\langle #1 \right\rangle}
\newcommand{\DOne}{\vec{\nabla}_{\!\vec{1}_1}}
\newcommand{\Done}{\vec{\nabla}_{\!\vec{1}}\,}
\newcommand{\Dk}{\vec{\nabla}_{\!\vec{k}}}
\newcommand{\f}[1]{f^{\texttt{#1}}}
\newcommand{\g}[1]{g^{(\texttt{#1})}}
\newcommand{\ff}[2]{f^{\texttt{#1}}_{\,#2}}
\newcommand{\ftwo}{f}
\renewcommand{\r}{\vec{r}}
\newcommand{\rp}{\vec{r}\,'}

\newcommand{\rhoss}[1]{\rho_{ij}^{\texttt{#1}}}
\newcommand{\rhos}[1]{\rho^{\texttt{#1}}}
\newcommand{\psis}[1]{\psi^{\texttt{#1}}}
\newcommand{\dss}[1]{d_{ij}^{\texttt{#1}}}
\newcommand{\ra}{\r_i}
\newcommand{\rb}{\r_j}
\newcommand{\thab}{\theta}
\newcommand{\psiF}{\widetilde{\psi}}
\newcommand{\fF}{\widetilde{f}}
\newcommand{\dtF}{\widetilde{d}^t_{ij}}

\begin{document}

\preprint{cond-mat/XXX}

\title{%
Statistical approach to dislocation dynamics: From dislocation
correlations to a multiple-slip continuum plasticity theory}

\author{Surachate Limkumnerd}
\email{s.limkumnerd@rug.nl}
\author{Erik Van der Giessen}
\email{E.van.der.Giessen@rug.nl}
\affiliation{%
Zernike Institute for Advanced Materials, Nijenborgh 4,
University of Groningen, 9747 AG Groningen, The Netherlands
}

\date{\today}

\begin{abstract}
Due to recent successes of a statistical-based nonlocal continuum
crystal plasticity theory for single-glide in explaining various
aspects such as dislocation patterning and size-dependent plasticity,
several attempts have been made to extend the theory to describe
crystals with multiple slip systems using ad-hoc assumptions. We
present here a mesoscale continuum theory of plasticity for multiple
slip systems of parallel edge dislocations. We begin by constructing
the Bogolyubov--Born--Green--Yvon--Kirkwood (BBGYK) integral equations relating
different orders of dislocation correlation functions in a grand
canonical ensemble. Approximate pair correlation functions are
obtained for single-slip systems with two types of dislocations and,
subsequently, for general multiple-slip systems of both charges. The effect of the correlations manifests itself in the form of an entropic force in addition to the external stress and the self-consistent internal stress. Comparisons with a previous multiple-slip theory based on phenomenological considerations shall be discussed.
\end{abstract}

\pacs{91.60.Ed,91.60.Dc}

\keywords{nonlocal plasticity; dislocations; pair correlations; BBGYK}

\maketitle

\section{Introduction}
Statistical mechanics provides an optimal framework and various tools for studying emergent phenomena from a complex conglomerate of bodies---may they be molecules of gases, polymer chains of rubber, or crystalline defects. The use of correlation functions in analysing two-dimensional solids and their defects has been proven very successful in the past. For example, Mermin showed that two-dimensional crystals do not have conventional long-range order, but can have ``directional long-range order.''\cite{Merm68} Nelson et al. applied the technique to explain dislocation-assisted melting in two dimensions.\cite{Nels78,NelsHalp78} Over a decade ago, Groma proposed a theory to describe dislocations and their motions using distribution functions and probability arguments.\cite{Grom97} Unlike the existing continuum theories at the time,%
\footnote{For a summary of various continuum theories, see, e.g., Ref.~\onlinecite{GhonBussKiouHuan03} and references therein.} the new formalism was physically motivated and incorporated correctly the long-range nature of dislocation interactions. Several variations of this work---all of which reduce to the same two-dimensional theory---also exist for three dimensional dislocation systems.\cite{ElAz00,LimkSeth06,LimkSeth07b,RoyAcha05}

Although having laid out the foundation for possible
interactions of many-dislocation configurations, Groma's pioneering work did not investigate these
correlated effects in details. Zaiser~et~al. considered explicitly the
evolution of dislocation correlations by extending Groma's theory for systems
of single-slip, parallel edge dislocations.\cite{ZaisMiguGrom01} They were
able to qualitatively obtain the correct scaling behavior of the evolution
equations for both single and pair correlation densities, and explained some
general properties of these functions. Their formulation, however, was limited
to only one active slip system and the analytical forms of pair correlation functions
were not derived. In a later work, Groma et al. included the influence of
dislocation correlations in the form of a local back
stress.\cite{GromCsikZais03} Yefimov~et~al. connected this statistical
description to a continuum crystal plasticity theory and applied this to
various boundary value problems.\cite{YefiGromGies04,YefiGromGies04b}
While the theory successfully captured most features observed in discrete
dislocation simulations, its ad-hoc extension to multiple slip systems failed to explain size effects in single crystal thin films.\cite{YefiGies05b} The main goals of this paper are: (1) to correctly describe and obtain analytical expressions for dislocation pair correlations, and (2) to systematically generalize the approach of Groma et al. to multiple slip systems.

We begin, in Sec.~\ref{S:Definitions}, by introducing ensembles of
dislocations and deriving the partition function for multiple slip
systems. The $n^\text{th}$-order dislocation densities and dislocation
correlation functions are subsequently defined. We construct the
Bogolyubov--Born--Green--Yvon--Kirkwood (BBGYK) integral equations in
Sec.~\ref{S:BBGYK}. These equations link correlation functions of order $n$ to
those of order $n\!+\!1$ (a technique generally used in the study of dense gases and fluids). The integral equations are expanded in powers of interaction strength (the ratio between the interaction energy and `thermal' energy). We then obtain a set of approximate integral equations for pair ($n=2$) correlation functions after applying a closure approximation to truncate the series. These equations are valid regardless of the form of the interaction potential, and thus are applicable to other systems, provided that this pair interaction vanishes at a large distance.

By appealing to Peach--Koehler interaction, analytical expressions for pair
dislocation densities for single and multiple slip systems are derived in
Sec.~\ref{S:SingleSlip} and Sec.~\ref{S:MultiSlip} respectively. Our
single-slip solution agrees with the result from the study of induced
geometrically necessary dislocations (GND) in terms of a single pinned
dislocation by means of a variational approach.\cite{GromGyorKocs06}
The dislocation spacing $1/\sqrt{\rho}$ emerges as a natural lengthscale in this
formulation in accordance with the scaling study by
Zaiser~et~al.\cite{ZaisMiguGrom01} Our analysis further shows long-range
attractive correlations when more than one slip system are present, confirming the absence of dislocation patterning in single glide systems as observed in many discrete dislocation simulations\cite{BenzBrecNeed04,BenzBrecNeed05,FourSala96,GomeDeviKubi06,GromBako00,GromPawl93PMA,GromPawl93MSEA,GullHart93} and explained in a recent three-dimensional continuum plasticity theory.\cite{LimkSeth06,LimkSeth07b}

In Sec.~\ref{S:EvolutionLaw}, we write down the transport equations for both
total dislocation densities and GND densities
on each slip system under the influence of Peach--Koehler forces from both
single and pair dislocation correlations. While the former gives a
self-consistent, long-range internal stress contribution, the latter exerts
an additional short-range, entropic force due to a deviation away from a
preferred dislocation arrangement in the form of a back stress. The
formulation is a direct extension of the work by Groma and
Zaiser\cite{Grom97,ZaisMiguGrom01,GromCsikZais03} for crystals with one active
slip system. Using knowledge of the pair correlation functions, we obtain a complete description of the back stress as a function of slip orientations---which previously had been incorporated using ad-hoc phenomenological considerations in the multiple-slip theory.\cite{YefiGies05b,YefiGies05}

Finally in Sec.~\ref{S:Comparison}, we contrast our theory with the multiple-slip theory of Yefimov~et~al.\cite{YefiGies05,YefiGies05b} While both theories propose that interactions among slip systems depend solely on relative angles of slip orientations, the functional forms are different. We attribute the failure of the earlier theory in explaining size effects in single crystal thin films partly to this difference and partly to the treatment of dislocation nucleation in the theory.

\section{Definitions of the basic quantities}\label{S:Definitions}

Consider a system containing $r$ species of dislocations and denote
the coordinate of the $i^\text{th}$ dislocation of species $s$ by
$\vec{i}_s$. The dislocation configuration $\set{N}$ is the set of the
coordinates of all dislocations, where $\system{N} \equiv (N_1,N_2;
N_3,N_4; \ldots; N_{r-1},N_{r})$ denotes the ``collection'' of
dislocations of type $s$. In this convention, odd and even slots
respectively contain plus and minus dislocations on distinct slip
systems.\footnote{Note that this already implies that the analysis
applies only to two-dimensional systems of dislocations.} We introduce the notation $\setplusone{N}{s}$ to denote the
addition of an extra dislocation of species $s$ to $\set{N}$, while
similarly a configuration $\set{N}$ with coordinates of $\system{n}$
removed is indicated by $\mysetminus$. 

The interacting Hamiltonian $U_\system{N}$ of the system can be written as the sum of potentials $u(\vec{i}_{s_1}-\vec{j}_{s_2})$ of all pairs of dislocations
\begin{equation}
  U_\system{N}(\set{N}) = \sum_{s_1 \le s_2}\sum_{i\le j} u(\vec{i}_{s_1}-\vec{j}_{s_2})\,.
\end{equation}
We can define a canonical partition function of configuration $\system{N}$ by
\begin{equation}\label{E:Zdef}
  \Z_\system{N} \equiv \int \exp{-U_\system{N}/\kB T} d\set{N}\,,
\end{equation}
where the integrations are taken over the ``volume'' measure $d\set{N} \equiv \prod_{s=1}^r d^2\vec{1}_s d^2\vec{2}_s \cdots d^2\vec{N}_s$ of the dislocation configuration at $\set{N}$.
Consider the coordinates of a particular set $\set{n}$, the probability of observing the configuration $\system{n}$ in $d\set{n}$ about the points in $\set{n}$ irrespective of the remaining collecttion $\system{N}-\system{n}$ is
\begin{equation}
  P_\system{N}^\order{n}(\set{n})\,d\set{n} = \frac{d\set{n}}{\Z_\system{N}} \!\int\! \exp{-U_\system{N}(\set{N})/\kB T} d\!\mysetminus,
\end{equation}
where $\int P_\system{N}^\order{n}(\set{n})\,d\set{n} = 1$.
The probability density of observing \emph{any} statistically equivalent possible collection $\system{n}$ within the volumes $d\set{n}$ about the points $\set{n}$ is therefore
\begin{equation}
  \rho_\system{N}^\order{n}(\set{n}) = \prod_{s=1}^r \frac{N_s!}{(N_s-n_s)!}\, P_\system{N}^\order{n}(\set{n})\,.
\end{equation}
By using Boltzmann distribution, we assume that our system is ergodic,
and thermal equilibrium exists and can be reached. System of
dislocations which drift along the local force (thus implying glide to
be accompanied by some amount of
climb) subject to thermal noise would certainly fit the criterion.

Consider now an \emph{open} system
(which could be realized, say, by allowing for nucleation and annihilation of dislocations as the system relaxes)%
; a grand canonical partition function is given by
\begin{equation}
  \Xi = \sum_{\system{N}\ge \system{0}} \prod_{s=1}^r \frac{z_s^{N_s}}{N_s!}\,\Z_\system{N}\,,
\end{equation}
where $z_s$ is the activity of species $s$. The prefactor arises from integrating away the momentum degrees of freedom in the Hamiltonian which are irrelevant to this problem. The probability $\mathcal{P}$ of the occurence of configuration $\system{N}$ in the open system is therefore
\begin{equation}\label{E:prob}
  \mathcal{P}_\system{N} = \prod_{s=1}^r \frac{z_s^{N_s}}{N_s!}\,\frac{\Z_\system{N}}{\Xi}\,.
\end{equation}
Finally, the probability density of observing \emph{any} $n_1$ dislocations of species 1, $n_2$ dislocations of species 2, etc., (\emph{any} collection $\system{n}$) in $d\set{n}$ at $\set{n}$ is
\begin{equation}\label{E:rhon}
  \rho^\order{n}(\set{n}) = \sum_{\system{N}\ge \system{n}} \mathcal{P}_\system{N}\, \rho^\order{n}_\system{N}(\set{n})\,.
\end{equation}
The summation is taken over all collections $\system{N}$ greater than or equal to $\system{n}$, i.e., for all $N_1\ge n_1$, $N_2\ge n_2$, etc.
We take Eq.~(\ref{E:rhon}) as the \emph{definition} of dislocation density of order $(\system{n})$. Explicitly we have
\begin{multline}\label{E:rho}
  \rho^\order{n}(\set{n}) = \frac{1}{\Xi} \sum_{\system{N}\ge \system{n}} \Big[\prod_{s=1}^r \frac{z_s^{N_s}}{(N_s-n_s)!} \Big] \\
\int \exp{-U_\system{N}(\set{N})/\kB T} d\!\mysetminus
\end{multline}
This definition of an $(\system{n})^\text{th}$-order dislocation density is equivalent to the ones used by Groma~\cite{Grom97} and Zaiser~\cite{ZaisMiguGrom01} in the realization of an open system.%
\footnote{%
One distinction due to the choice of an emsemble type can be seen from the normalization condition. In the grand canonical ensemble, according to Eqs. (\ref{E:prob}) and (\ref{E:rhon}),
$\int \rho^\order{n}(\set{n})\,d\set{n} = \sum_{\system{N}\ge \system{n}} \mathcal{P}_\system{N} \Big[\prod_{s=1}^r \frac{N_s!}{(N_s-n_s)!} \Big] = \avg{\prod_{s=1}^r \frac{N_s!}{(N_s-n_s)!} }$,
where $\avg{\cdot}$ denotes an average over all statistically equivalent ensembles.
In particular, the density of a single dislocation of species $s$ in a system with no external shear, $\rho^{(1)}(\vec{1}_s)$ is independent of $\vec{1}_s$, thus, $\int \rho^{(1)}(\vec{1}_s) \, d^2\vec{1}_s = \rho^{(1)}_s \,A = \avg{N_s}$. In other words $\rho^{(1)} = \avg{N_s}/A$ depends on the \emph{average} number of dislocation of species $s$. If one were to carry a similar analysis in a canonical ensemble where the number of dislocations of each species is fixed, $\rho^{(1)}_s$ would have to be replaced by $N_s/A$, where $N_s$ is fixed. For higher order density, the expression becomes quite cumbersome. For example, $\rho^{(2)}_{ss'} = N_s N_s'/A^2$ for $s\ne s'$, while for $s = s'$, it is $N_s(N_s-1)/A^2$. In thermodynamic limit ($N_s\rightarrow \infty$ and $A\rightarrow \infty$ while keeping the ratio fixed), these two expressions reduce to the same thing. In this sense, it is cleaner to work in the grand canonical ensemble.}
Finally, we define the $\order{n}^\text{th}$-order correlation function $g^\order{n}(\set{n})$ through
\begin{equation}\label{E:g}
  \rho^\order{n}(\set{n}) = \left[\prod_{s=1}^r \rho^{(1)}(\vec{1}_s)\rho^{(1)}(\vec{2}_s)\cdots \rho^{(1)}(\vec{n}_s)\right] g^\order{n}(\set{n})\,.
\end{equation}

\section{Derivation of the BBGYK integral equations}\label{S:BBGYK}
The Bogolyubov--Born--Green--Yvon--Kirkwood integral equations first appeared in the study of classical fluids with a total potential energy given by the sum of pair interactions.\cite{Kirk35,Yvon35,BornGree49,Gree52} They provide a set of relations between distribution functions of fluid density at different orders. Here we extend the BBGYK formalism to include the non-central interactions of dislocations in a multicomponent system.\cite{Fish64,Hill56} We proceed in three steps: (1) take a derivative of the $\order{n}^\text{th}$-order dislocation density with respect to the coordinate of one particle of the interested species; (2) express the result in terms of the next higher order densities; and (3) convert the integral equations of densities into those of correlation functions.

Differentiating $\rho^\order{n}(\set{n})$ as expressed in
Eq.~(\ref{E:rho}) with respect to dislocation 1 of species 1 located
at $\vec{1}_1$ we find
\begin{multline}\label{E:step1}
  \DOne \rho^\order{n}(\set{n}) = -\frac{1}{\Xi} \sum_{\system{N}\ge \system{n}} \Big[\prod_{s=1}^r \frac{z_s^{N_s}}{(N_s-n_s)!} \Big] \\
\int \exp{-\bar{U}_\system{N}(\set{N})} \DOne \bar{U}_\system{N}(\set{N})\, d\!\mysetminus\,,
\end{multline}
where we absorb $1/\kB T$ into the definition
$\bar{U}_\system{N}:=U_\system/\kB T$.
The derivative of the potential can be separated into two parts:
\begin{equation}\label{E:DofPotential}
  \DOne \bar{U}_\system{N} = \underset{(i,s)\ne (1,1)}{\sum_{s=1}^r\sum_{i=1}^{n_s}} \DOne \bar{u}(\vec{1}_1-\vec{i}_s) + \sum_{s=1}^r\sum_{i=n_s+1}^{N_s} \DOne \bar{u}(\vec{1}_1-\vec{i}_s)
\end{equation}
Direct substitution of Eq.~(\ref{E:DofPotential}) into the integrand of
Eq.~(\ref{E:step1}) splits the expression into two integrals $I_1$ and
$I_2$. Notice in the first integral that the derivative of the potential does
not depend on coordinates $\mysetminus$, and thus can be taken out of the
integral, yielding
\begin{equation}\label{E:I1}
I_1 = - \rho^\order{n}(\set{n}) \underset{(i,s)\ne (1,1)}{\sum_{s=1}^r\sum_{i=1}^{n_s}} \DOne \bar{u}(\vec{1}_1-\vec{i}_s)
\end{equation}
with the aid of Eq.~(\ref{E:rho}). The second integral $I_2$ requires a little more work:
\begin{multline}\label{E:step2}
  I_2 = -\frac{1}{\Xi} \sum_{s=1}^r \sum_{\system{N}\ge \system{n}} \Big[\prod_{s'=1}^r \frac{z_{s'}^{N_{s'}}}{(N_{s'}-n_{s'})!} \Big] \\
\int \! \sum_{i=n_s+1}^{N_s} \DOne \bar{u}(\vec{1}_1-\vec{i}_s)\, \exp{-\bar{U}_\system{N}(\set{N})} \, d\!\mysetminus
\end{multline}
The expression involves integrating $\vec{i}_s$ over the sample size. Since each integral over $\vec{i}_s$ between $n_s+1 \le i \le N_s$ is equivalent in infinite space, the summation therefore gives a factor of $(N_s-n_s)$. The remaining integrals over all other dislocation coordinates are unaffected. \begin{widetext}
Eq.~(\ref{E:step2}) thus becomes
\begin{equation}\label{E:I2}
\begin{split}
  I_2 &= -\sum_{s=1}^r \frac{1}{\Xi} \sum_{\system{N}\ge \system{n}} \Big[\prod_{s'=1}^r \frac{z_{s'}^{N_{s'}}}{(N_{s'}-n_{s'})!} \Big] (N_s-n_s) \\
&\qquad \int \DOne \bar u(\vec{1}_1-\overrightarrow{(n_s+1)}_s) \Big\{ \int \exp{-\bar{U}_\system{N}(\set{N})} \, d\!\left[\mysetminus \!\setminus\!\{\overrightarrow{(n_s \!+\!1)}_s\}\right] \Big\} \,d^2\overrightarrow{(n_s\!+\!1)}_s \\
&= -\sum_{s=1}^r \int \DOne \bar u(\vec{1}_1-\overrightarrow{(n_s+1)}_s) \bigg\{ \frac{1}{\Xi} \sum_{\system{N}\ge \system{n}} \Big[\prod_{s'=1}^r \frac{z_{s'}^{N_{s'}}(N_s-n_s)}{(N_{s'}-n_{s'})!} \Big] \\
&\quad\qquad\qquad\qquad\qquad\qquad\qquad \int \exp{-\bar{U}_\system{N}(\set{N})} \, d\!\left[\mysetminus \!\setminus\!\{\overrightarrow{(n_s \!+\!1)}_s\}\right] \bigg\} \,d^2\overrightarrow{(n_s\!+\!1)}_s \\
&= -\sum_{s=1}^r \int \DOne \bar u(\vec{1}_1-\overrightarrow{(n_s+1)}_s)\, \rho^\orderplusone{n}{s}(\setplusone{n}{s})\,d^2\overrightarrow{(n_s\!+\!1)}_s
\end{split}
\end{equation}
The symbol $d\!\left[\mysetminus \!\setminus\!\{\overrightarrow{(n_s \!+\!1)}_s\}\right]$ represents the volume measure of $\mysetminus$ \emph{without} $d^2\overrightarrow{(n_s\!+\!1)}_s$.
Collecting both $I_1$ and $I_2$ from Eq.~(\ref{E:I1}) and Eq.~(\ref{E:I2}), we arrive at the BBGYK equations for the $\order{n}^\text{th}$-order dislocation density:
\begin{equation}\label{E:BBGYKrho}
  \DOne \rho^\order{n}(\set{n}) = - \rho^\order{n}(\set{n}) \!\!\!\sum_{(s,i)\ne (1,1)}^{(r,n_s)}\!\!\!\!\! \DOne \bar{u}(\vec{1}_1-\vec{i}_s)
-\sum_{s=1}^r \int \DOne \bar u(\vec{1}_1-\overrightarrow{(n_s\!+\!1)}_s)\, \rho^\orderplusone{n}{s}(\setplusone{n}{s})\,d^2\overrightarrow{(n_s\!+\!1)}_s
\end{equation}

One can obtain a series of integro-differential equations for the correlation
functions $g^\order{n}$ from Eq.~(\ref{E:BBGYKrho}) by expanding out
$\rho^\order{n}(\set{n})$ using Eq.~(\ref{E:g}). All but two of the single
dislocation densities on the left and right-hand sides of the equality cancel which results in
\begin{multline}\label{E:BBGYKgfirst}
  \DOne\! \left[\rho(\vec{1}_1) g^\order{n}(\set{n})\right] = - \rho(\vec{1}_1) g^\order{n}(\set{n}) \!\!\!\sum_{(s,i)\ne (1,1)}^{(r,n_s)}\!\!\!\!\! \DOne \bar{u}(\vec{1}_1-\vec{i}_s) \\
-\rho(\vec{1}_1) \sum_{s=1}^r \int \DOne \bar u(\vec{1}_1-\overrightarrow{(n_s\!+\!1)}_s)\, \rho(\overrightarrow{(n_s\!+\!1)}_s)g^\orderplusone{n}{s}(\setplusone{n}{s})\,d^2\overrightarrow{(n_s\!+\!1)}_s\,.
\end{multline}
\end{widetext}
The first order densities $\rho(\vec{1})$ that plague the expression can be
removed by first using the product rule to the left-hand side (LHS), then dividing both sides by $\rho(\vec{1})$. The LHS becomes
\begin{equation*}
  \text{LHS} = \DOne g^\order{n}(\set{n}) + g^\order{n}(\set{n})\frac{\DOne\rho(\vec{1}_1)}{\rho(\vec{1}_1)}.
\end{equation*}
The ratio of the derivative of the first-order density with itself can be rewritten using Eq.~(\ref{E:BBGYKrho}) specialized to first order, giving
\begin{equation*}
  \frac{\DOne\rho(\vec{1}_1)}{\rho(\vec{1}_1)} = -\sum_{s=1}^r \int \DOne \bar u(\vec{1}_1-\vec{\xi}_s) \rho(\vec{\xi}_s) g^{(2)}(\vec{1}_1,\vec{\xi}_s)\,d^2\vec{\xi}_s\,,
\end{equation*}
where $\vec{\xi}_s \equiv \overrightarrow{(n_s\!+\!1)}_s$ is the position of
the $(n_s\!+\!1)^\text{th}$ dislocation of species $s$, and
$g^{(2)}(\vec{1}_1,\vec{\xi}_s)$ represents the pair correlation function
between the first dislocation of species 1 at $\vec{1}_1$ and the
$(n_s+1)^\text{th}$ dislocation of species $s$ at $\vec{\xi}_s$. This
expression could be incorporated seamlessly into the right-hand side of Eq.~(\ref{E:BBGYKgfirst}). The final result is%
\footnote{In the presence of an external \emph{conservative} force, it can be shown that both (\ref{E:BBGYKrho}) and (\ref{E:BBGYKg}) remain valid provided that an additional term representing the applied external force, $\vec{F}(\vec{1}_1) \equiv -(1/\kB T)\DOne \Phi(\vec{1}_1)$ generated by the external potential $\Phi(\vec{1}_1)$ which acts on $\vec{1}_1$, is added to their RHS. Qualitatively speaking, the original expression is nothing but the sum of all the Peach--Koehler interactions on the dislocation at $\vec{1}_1$ due to all other dislocations in the collection $\system{n}$.}
\begin{multline}\label{E:BBGYKg}
\DOne g^\order{n}(\set{n}) = - g^\order{n}(\set{n}) \!\!\!\sum_{(s,i)\ne (1,1)}^{(r,n_s)}\!\!\!\!\! \DOne \bar{u}(\vec{1}_1-\vec{i}_s) \\
- \sum_{s=1}^r \int \DOne \bar u(\vec{1}_1-\vec{\xi}_s)\, \rho(\vec{\xi}_s) \\ \times\left[g^\orderplusone{n}{s}(\setplusone{n}{s})- g^\order{n}(\set{n})g^{(2)}(\vec{1}_1,\vec{\xi}_s) \right] d^2\vec{\xi}_s\,.
\end{multline}

For the remainder of this paper, we shall restrict our attention to the Peach--Koehler interaction. Recall that the interaction energy between two parallel edge dislocations of length $L$ (over thermal energy $\kB T$) in an infinite medium is~\cite{HirtLoth82}
\begin{equation}\label{E:energy}
  \bar u(\vec{i}_s-\vec{j}_{s'}) = -\Gamma\, \psi(\vec{i}_s-\vec{j}_{s'})
\end{equation}
where $\Gamma \equiv \dfrac{\mu b^2 L}{2\pi (1-\nu) \kB T}$, and
\begin{multline}\label{E:IntEnergy}
\psi(\vec{i}_s,\vec{j}_{s'}) \equiv \bigg[ (\hat{m}_{\vec{i}_s}\cdot \hat{m}_{\vec{j}_{s'}}) \ln\!\left(| \vec{i}_s - \vec{j}_{s'} |\right) \\
+ \frac{\left(\hat{m}_{\vec{i}_s}\cdot (\vec{i}_s - \vec{j}_{s'})\right) \left(\hat{m}_{\vec{j}_{s'}}\cdot (\vec{i}_s - \vec{j}_{s'})\right)}{|\vec{i}_s - \vec{j}_{s'}|^2} \bigg].
\end{multline}
Here $\hat{m}_{\vec{i}_s}$ denotes the slip-plane normal of species $s$. The relative strength $\Gamma$ represents the ratio between 
dislocation interaction energy versus the system's thermal
energy. Note that the latter originates from the use of Boltzmann distribution in Eq.~(\ref{E:Zdef})
to describe the equilibrium configuration of systems with thermal noise. It was pointed out by Groma~et~al.\cite{GromGyorKocs06} that, in systems where dislocations are confined to their slip planes, the glide constraint acts as an effective temperature preventing the systems to relax by means of dislocation annihilation.
Seen in this light, the temperature $T$ in this theory is not
physical temperature but a fictive temperature associated with the
disorder in dislocation distributions.\footnote{There are some systems where climb is typical and $\Gamma$ is naturally small such as dislocations in vortex lattices of type II superconductor where the values of elastic moduli can be small at suitably applied magnetic field.\cite{BlatFeigGeshLark94} By modifying the form of the interaction potential, the present analysis can be carried over straightforwardly.}
As the dislocation configuration becomes more and more correlated,
$\Gamma$ becomes smaller.

For an explicit dependence on $\Gamma$ to use as an expansion coefficient, we rescale the distance by the square-root of the relative strength, $\sqrt{\Gamma}\,\vec{r} \mapsto \vec{r}$. Eq.~(\ref{E:BBGYKg}) specialized to second order gives
\begin{multline}\label{E:gsecondorder}
\Done g^{(2)}(\vec{1},\vec{2}) = \Gamma\, g^{(2)}(\vec{1},\vec{2}) \Done \psi(\vec{1},\vec{2}) \\
+ \sum_{s=1}^r \int \Done \psi(\vec{1},\vec{3}_s)\, \rho(\vec{3}_s) \\ \times\left[g^{(3)}(\vec{1},\vec{2},\vec{3}_s)- g^{(2)}(\vec{1},\vec{2})g^{(2)}(\vec{1},\vec{3}_s) \right] d^2\vec{3}_s\,.
\end{multline}
Here we have simplified the notation even further by suppressing all
irrelevant subscripts: vectors $\vec{1}$ and $\vec{2}$ simply denote the positions of dislocations 1 and 2 with their corresponding species. The summation $\sum_r$ is taken over all $s$ species present in the system.

We proceed by assuming that the correlation functions have the following forms:
\begin{subequations}\label{E:g23}
\begin{align}
\begin{split}\label{E:g2}
  g^{(2)}(\vec{1},\vec{2}) &= 1 + \Gamma\,f^{(2)}(\vec{1},\vec{2})\,,
\end{split} \\
\begin{split}\label{E:g3}
  g^{(3)}(\vec{1},\vec{2},\vec{3}) &= 1 + \Gamma\left[f^{(2)}(\vec{1},\vec{2}) + f^{(2)}(\vec{1},\vec{3}) + f^{(2)}(\vec{2},\vec{3}) \right] \\
&\qquad + \Gamma^2\, f^{(3)}(\vec{1},\vec{2},\vec{3})\,,
\end{split}%
\end{align}
\end{subequations}
for any vectors $\vec{1}$, $\vec{2}$, and $\vec{3}$.
The functions $f^{(2)}(\vec{1},\vec{2})$ and $f^{(3)}(\vec{1},\vec{2},\vec{3})$ should asympotically vanish along the boundaries of the sample, or as $|\vec{1}-\vec{2}|, |\vec{1}-\vec{3}|, |\vec{2}-\vec{3}| \rightarrow \infty$ for an infinite system. Note in particular that
\begin{multline}\label{E:gsubtract}
g^{(3)}(\vec{1},\vec{2},\vec{3})- g^{(2)}(\vec{1},\vec{2})g^{(2)}(\vec{1},\vec{3}) = \Gamma\,f^{(2)}(\vec{2},\vec{3}) \\
+ \Gamma^2\!\left[f^{(3)}(\vec{1},\vec{2},\vec{3}) - f^{(2)}(\vec{1},\vec{2})f^{(2)}(\vec{1},\vec{3})\right]\!.
\end{multline}

So far no approximation has been made. The Eqs.~(\ref{E:g23}) governing the
second--order correlations naturally involve the third--order
correlations. To systematically close the chain at the second order,
we substitute Eqs.~(\ref{E:g23}) and (\ref{E:gsubtract}) into
Eq.~(\ref{E:gsecondorder}) to produce a set of integro--differential
equations of $f^{(2)}$ and $f^{(3)}$ for each power of $\Gamma$. This
technique was introduced by Bogolyubov in the study of correlations in
Coulomb interactions~\cite{Bogo46} and has since been widely used in both high energy and condensed matter communities in renormalization group theory.

The equation of power $\Gamma^0$ gives an identity. After integrating away $\Done$ because $f^{(2)}$
\emph{and} $\psi$ vanish along a boundary, the equation of power $\Gamma$ becomes,
\begin{equation}\label{E:BBGYKf}
f^{(2)}(\vec{1},\vec{2}) = \psi(\vec{1},\vec{2})
+ \sum_{s=1}^r \int \psi(\vec{1},\vec{3}_s) \rho(\vec{3}_s) f^{(2)}(\vec{2},\vec{3}_s)\, d^2\vec{3}_s
\end{equation}
This equation is the key result of the analysis. In the following sections, we
shall use it to obtain dislocation pair correlation functions for systems with
one (Sec.~\ref{S:SingleSlip}) and many (Sec.~\ref{S:MultiSlip}) active slip systems.

\section{Pair correlation functions for single slip}\label{S:SingleSlip}
To illustrate the use of Eq.~(\ref{E:BBGYKf}), we first apply it to the
case of one slip system containing two species of dislocations (denoted $+$ and $-$). 
According to Eq.~\ref{E:IntEnergy} valid for an infinite sample, $\psi(\vec{1},\vec{2}) = \psi(\vec{1}-\vec{2}) = \psi(\vec{2}-\vec{1})$ which implies that $f^{ab}(\vec{1},\vec{2}) = f^{ab}(\vec{1}-\vec{2}) = f^{ab}(\vec{2}-\vec{1})$. Without loss of generality, we can take the origin to be at $\vec{2}$ and thus, from (\ref{E:BBGYKf}), we obtain the following set of integral equations:
\begin{subequations}\label{E:BBGYKoneslip}
\begin{align}
\begin{split}
\f{++}(\r) &= \phantom{-}\psi_1(\r) + \int d^2\rp\, \psi_1(\r-\rp) \\ &\qquad\left[\rhos{+}(\rp)\f{++}(\rp) - \rhos{-}(\rp)\f{+-}(\rp)\right] \label{E:fpp}
\end{split}\\
\begin{split}
\f{+-}(\r) &= -\psi_1(\r) - \int d^2\rp\, \psi_1(\r-\rp) \\
&\qquad\left[\rhos{-}(\rp)\f{--}(\rp) - \rhos{+}(\rp)\f{-+}(\rp)\right] \label{E:fpm}
\end{split}\\
\begin{split}
\f{--}(\r) &= \phantom{-}\psi_1(\r) + \int d^2\rp\, \psi_1(\r-\rp) \\ &\qquad\left[\rhos{-}(\rp)\f{--}(\rp) - \rhos{+}(\rp)\f{-+}(\rp)\right] \label{E:fmm}
\end{split}\\
\begin{split}
\f{-+}(\r) &= -\psi_1(\r) - \int d^2\rp\, \psi_1(\r-\rp) \\
&\qquad\left[\rhos{+}(\rp)\f{++}(\rp) - \rhos{-}(\rp)\f{+-}(\rp)\right] \label{E:fmp}
\end{split}
\end{align}
\end{subequations}
In the current context, Eq.~(\ref{E:IntEnergy}) reduces to
\begin{equation}\label{E:InteractionPotential}
  \psi_1(\vec{r}) = \psi^{\texttt{++}}(\vec{r}) = -\psi^{\texttt{+-}}(\vec{r}) = \ln(|\vec{r}|) + \frac{y^2}{|\vec{r}|^2}\,,
\end{equation}
where we orient our $(x,y)$ coordinate system in such a way that the slip direction points along the ${x}$ direction. The minus signs in Eq.~(\ref{E:BBGYKoneslip}) arise from a sign difference in the interactions between plus--plus dislocations versus plus--minus dislocations as shown in Eq.~(\ref{E:InteractionPotential}).
By comparing Eq.~(\ref{E:fpp}) against (\ref{E:fmp}), and Eq.~(\ref{E:fpm}) against (\ref{E:fmm}), we find that
$\f{++}(\r) = -\f{-+}(\r)$ and $\f{+-}(\r) = -\f{--}(\r)$. These symmetries further imply that $\f{++}(\r) = \f{--}(\r)$. Finally we obtain
\begin{multline}\label{E:intEqn}
\f{++}(\r) = \psi_1(\r) \\
+ \int \psi_1(\r-\rp) \f{++}(\rp) \left[\rhos{+}(\rp) + \rhos{-}(\rp)\right] d^2\rp\,.
\end{multline}

Our general formulation in the previous section allows for spatial variation of an uncorrelated density $\rho(\r_s)$. Without externally applied force, $\rho(\r_s) = \avg{N_s}/A$ is constant in space. An analytical solution to Eq.~\ref{E:intEqn} can be obtained for constant $\rhos{+}$ and $\rhos{-}$.
The dimensionless nature of the interaction potential $\psi_1$ suggests a change of variable $\sqrt{\rhos{+}+\rhos{-}}\,\r \mapsto \r$ (note that $\rhos{+}$ and $\rhos{-}$ are always positive). The resulting dimensionless integral equation
\begin{equation}\label{E:fppdimless}
\f{++}(\r) = \psi_1(\r) + \int \psi_1(\r-\rp) \f{++}(\rp) d^2\rp
\end{equation}
can be solved directly by applying $\Delta^2 \equiv (\partial_x^2 +\partial_y^2)^2$ on both sides of the equation and using the identity
\begin{equation}
\Delta^2 \psi_1(\r) = 2\pi\Delta\delta(\r) +2\pi (\partial^2_y - \partial^2_x)\delta(\r) = 4\pi\partial^2_y\delta(\r).
\end{equation}
Eq.~(\ref{E:fppdimless}) then becomes
\begin{equation}
\Delta^2 \f{++} = 4\pi \partial^2_y\!\left[\f{++} + \delta(\r) \right]\,,
\end{equation}
whose explicit solution is
\begin{equation}\label{E:fsingleslip}
  \f{++} = \frac{y}{r}\sinh(\sqrt{\pi}y)K_1(\sqrt{\pi}r) - \cosh(\sqrt{\pi}y)K_0(\sqrt{\pi}r)\,,
\end{equation}
with $K_0(\cdot)$ and $K_1(\cdot)$ the zeroth and first order modified
Bessel functions of the second kind. With the aid of Eq.~(\ref{E:g2}), the correlation functions $\g{++} = \g{--}$ and $\g{+-} = \g{-+}$, correct to $\mathcal{O}(\Gamma^2)$, can be expressed in the original coordinates,
\begin{subequations}\label{E:goneanswer}
\begin{align}
\begin{split}
\g{++}(\r) = 1 + \Gamma\bigg[\frac{y}{r}\sinh(k_0 &y) K_1(k_0 r) \\ &- \cosh(k_0 y)K_0(k_0 r)\bigg], \label{E:goneanswer1}
\end{split}\\
\begin{split}
  \g{+-}(\r) = 1 - \Gamma\bigg[\frac{y}{r}\sinh(k_0 &y) K_1(k_0 r) \\ &- \cosh(k_0 y)K_0(k_0 r)\bigg], \label{E:goneanswer2}
\end{split}
\end{align}
\end{subequations}
where $k_0 \equiv \sqrt{\pi \Gamma (\rhos{+}+\rhos{-})}$ gives an inverse ``Debye radius'' of the dislocation cloud. The third order correlation functions correct up to $\mathcal{O}(\Gamma^2)$ follow straightforwardly from Eq.~(\ref{E:g3}).
The validity of Eq.~(\ref{E:goneanswer}) can be verified by comparing
$\g{++}(\r)-\g{+-}(\r)$ with the dislocation difference, or GND, field
$\kappa(\r)$ in Eq.~(15) of Ref.~\onlinecite{GromGyorKocs06}. 
In this latter work the
\emph{same} expression is obtained for the induced GND due to a single pinned
dislocation, which was interpreted by the authors as the pair correlation of dislocations in a relaxed system.

It is interesting to note that the pair correlation functions depend only on
the scaled space coordinate $\sqrt{\rho}\,\r$ ($\rho \equiv
\rhos{+}+\rhos{-}$ being the total dislocation density) in agreement with the scaling argument given by Zaiser~et~al.\cite{ZaisMiguGrom01} This dependence also holds in the multiple-slip case to be discussed in the next section.

\section{Pair correlation functions for multiple slip}\label{S:MultiSlip}
The procedure to obtain the correlation functions for a system with
multiple slips follows the same types of arguments and expansions as
those for single slip. We shall further develop the integral equation
(\ref{E:BBGYKf}) for a system of $N$ slip systems, each with two charges, and
subsequently give an explicit analytical solution for the pair correlation
function in the case where the difference in slip orientation angle between
adjacent slip planes is constant.

For an $N$-slip system with both types of charges, we have $4N^2$
coupled integral equations for different pairs of $\vec{1}$ and
$\vec{2}$ in Eq.~(\ref{E:BBGYKf}). To reduce the number of equations,
and essentially decouple them, some symmetry arguments can be employed. For an infinite system,
\begin{equation*}
  \psis{++}_{ij} = \psis{--}_{ij} = -\psis{+-}_{ij} = -\psis{-+}_{ij}\,, \quad\text{and}\quad \psi^{ab}_{\,ij} = \psi^{ab}_{\,ji}\,,
\end{equation*}
where the superscripts denote the charges of the first and second
dislocations, while the subscripts show the slip systems in which they
live. Eq.~(\ref{E:BBGYKf}) can be re-cast using the convolution operator $\ast$
and the symmetry of $\psi^{ab}_{\,ij}$ as
\begin{equation}\label{E:ftemp}
  f^{ab}_{\,ij} = \psi^{ab}_{\,ij} + \sum_{k=1}^N \psi^{a\texttt{+}}_{ik}\ast\left[\rhos{+}_k f^{b\texttt{+}}_{jk} - \rhos{-}_k f^{b\texttt{-}}_{jk} \right].
\end{equation}
By direct substitution of $+$ and $-$ into $a$ and $b$, it is immediate that $\ff{++}{ij}(\r) = -\ff{-+}{ij}(\r)$ and $\ff{--}{ij}(\r) = -\ff{+-}{ij}(\r)$, which further implies that
\begin{equation}\label{E:fsimplified}
  \ff{++}{ij} = \ff{--}{ij} = \psis{++}_{ij} + \sum_{k=1}^N \psis{++}_{ik}\ast\left[\rhos{+}_k \ff{++}{\!jk} + \rhos{-}_k \ff{--}{\!jk} \right].
\end{equation}
With this, Eq.~(\ref{E:ftemp}) reduces to
\begin{equation}\label{E:fn}
\ftwo_{ij} = \psi_{ij} + \sum_{k=1}^N \psi_{ik}\ast \big[\rho_k \ftwo_{jk}\big],
\end{equation}
where the superscripts have been omitted and $\rho_k \equiv
\rhos{+}_k+\rhos{-}_k$ is the total dislocation density of both types
on slip $k$. We thus effectively reduce the number of coupled
equations to $N^2$. Note also that because of $\psi_{ij} = \psi_{ji}$, there are only $N(N+1)/2$ independent $\psi_{ij}$'s.

As seen from the single-slip case, Eq.~(\ref{E:fn}) subjected to an arbitrary distribution of the local density $\rho_k(\r)$ cannot be solved analytically. For spatially independent $\rho_k$, however, these equations can be decoupled. Let $\lambda_k$ be the relative population of density in slip system $k$ relative to the total density $\rho$, i.e., $\rho_k = \lambda_k \rho$ where $\sum_{k=1}^N \lambda_k = 1$. We can then perform a change of variable $\sqrt{\rho}\,\r \mapsto
\r$ to absorb the $\rho$--dependence. In
addition, in Fourier space (indicated by a superposed $\sim$), a convolution
becomes a product. We can solve the Fourier-transform of (\ref{E:fn}) for $\fF_{ij}$ by essentially performing a matrix inversion on
\begin{equation}\label{E:fF}
  \psiF_{ij} = \sum_{m,n}(\delta_{im}\delta_{jn} - \lambda_n\psiF_{in}\delta_{jm})\fF_{mn}\,.
\end{equation}

The Fourier representation of $\psi_{ij}$ in Eq.~(\ref{E:IntEnergy}) can be expressed very simply in polar coordinates $(k,\phi_k)$,
\begin{equation}\label{E:psi}
  \psiF_{ij} = -\frac{4\pi}{k^2}\sin(\phi_k-\theta_i)\sin(\phi_k-\theta_j) = -\frac{4\pi}{k^4}\,(\hvect{m}_i\cdot\vect{k})(\hvect{m}_j\cdot\vect{k})
\end{equation}
where $\theta_i$ is the angle that slip plane $i$ makes with the ${x}$
axis (which can be chosen arbitrarily, so that $\theta_i=i\pi/N$).
Owing to the simple form of (\ref{E:psi}), the solution to (\ref{E:fF}) is\footnote{The form of the solution is not surprising; it suggests that the solution can be written as a sum of diagrams due to the expansion $1/(1-x) = 1+x+x^2+\ldots$, often encountered in a many-body theory.}
\begin{equation}\label{E:fsoln}
  \fF_{ij} = \frac{\psiF_{ij}/\lambda_j}{1-\sum_n\psiF_{nn}}
\end{equation}
where we have used $\sum_n\psiF_{in}\psiF_{nj} = \psiF_{ij}\sum_n\psiF_{nn}$. 
Eq.~(\ref{E:fsoln}) shall be used in the derivation of the evolution law for parallel edge dislocations in a multislip system in the next section.

\begin{figure}[htbp]
\centering
(a)

\includegraphics[width=0.4\textwidth]{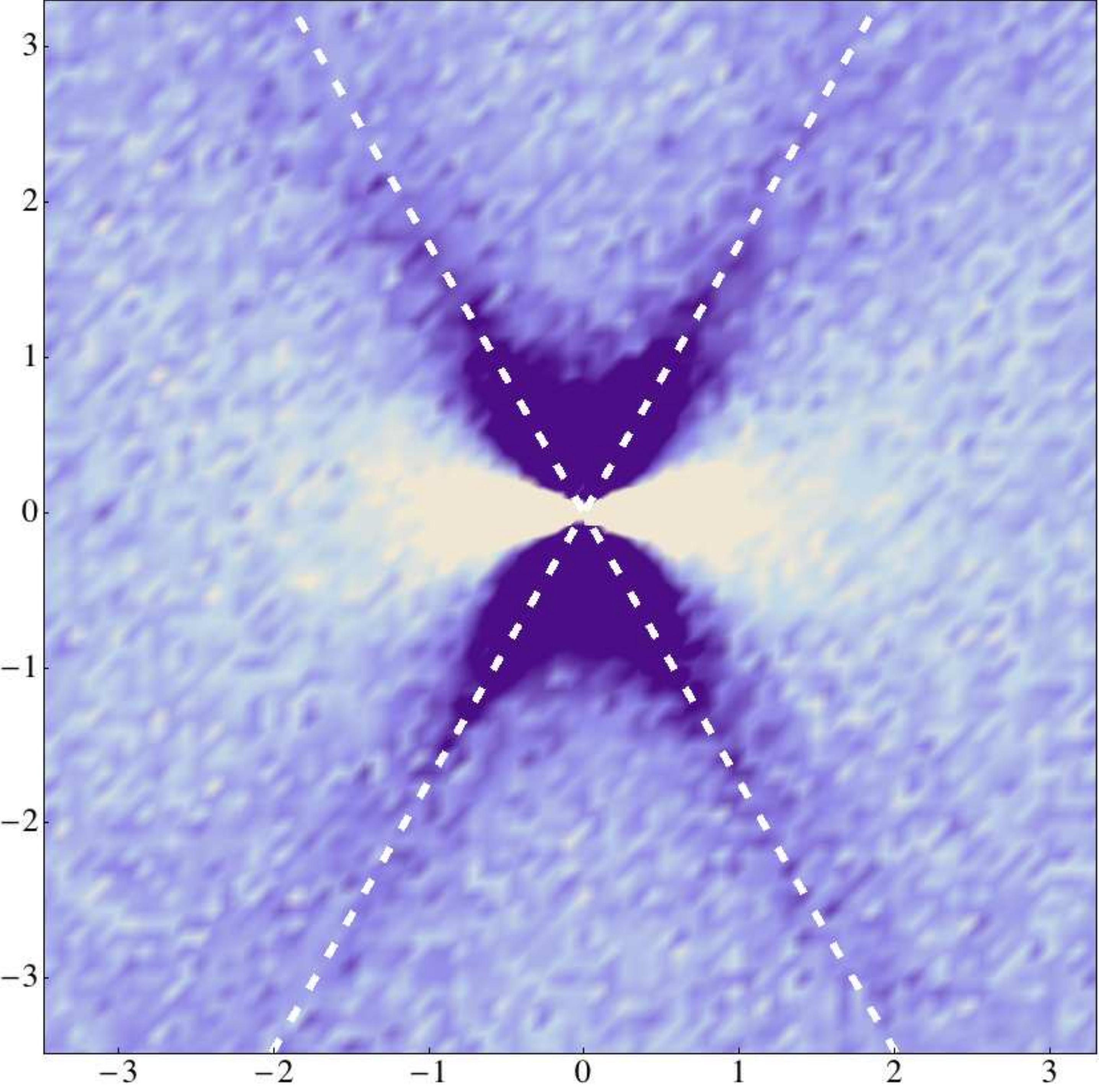}\\
(b)

\includegraphics[width=0.4\textwidth]{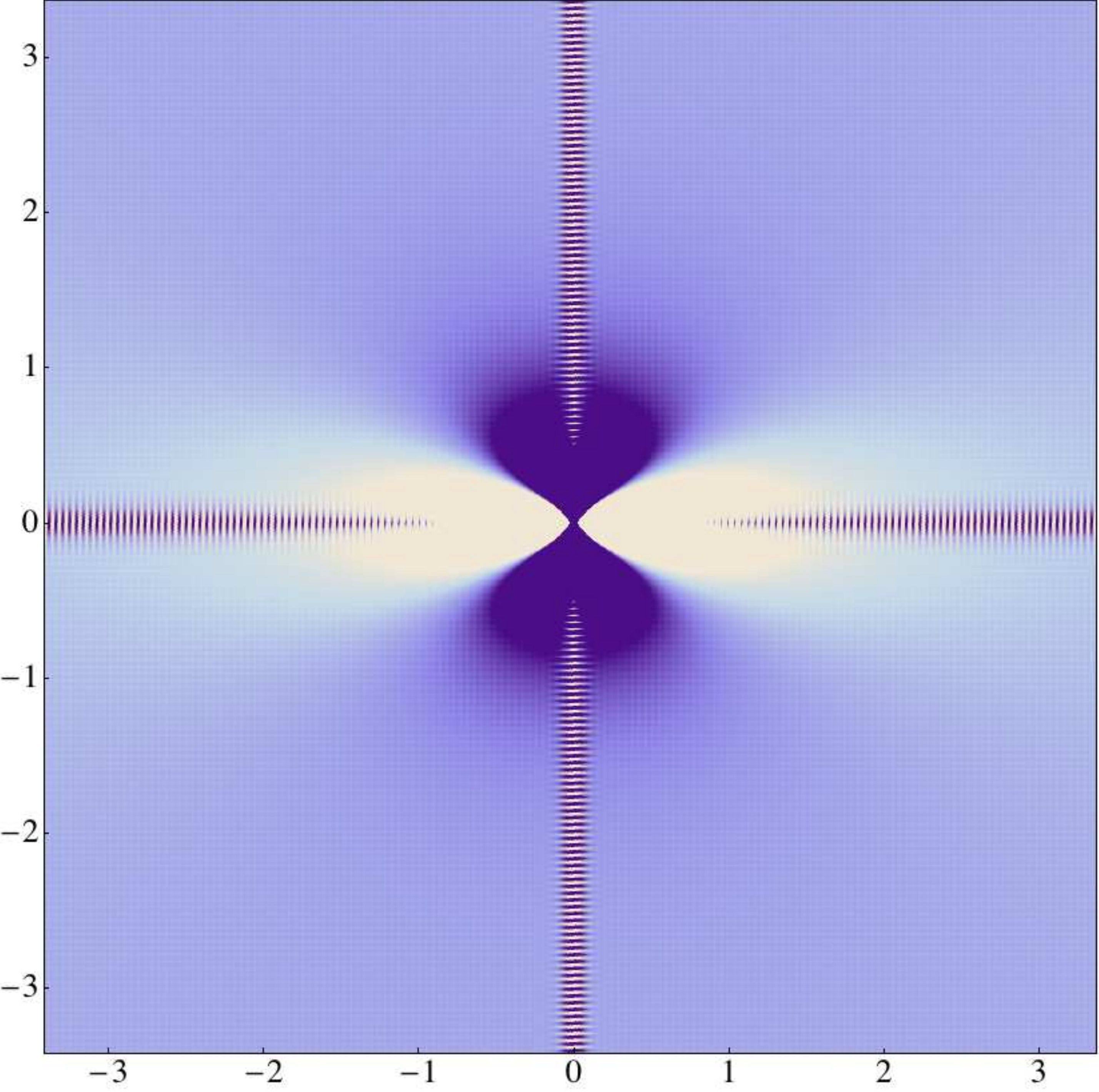}\\
\caption{(a) Discrete dislocation result and (b) theoretical
prediction of the correlation function $\ff{++}{12}$ between plus dislocations on $60^\circ$ and $120^\circ$ slip systems. Values increase towards brighter regions. Coordinates are measured in units of $1/\sqrt{\rho}$. Dashed lines indicate the two slip directions where the plus-plus anti-correlation is underpredicted due to the glide constraint of the discrete dislocation simulations. The fitting parameter due to rescalings of length was found to be $k_0 \simeq
22\sqrt{\rho}$. }
\label{Fig:DensityPlot}
\end{figure}
\begin{figure}[htbp]
\begin{center}
\psfrag{d}{$\ff{++}{12}$}
\psfrag{xinrho}{$x~({1}/{\sqrt{\rho}})$}
\includegraphics[width=0.47\textwidth]{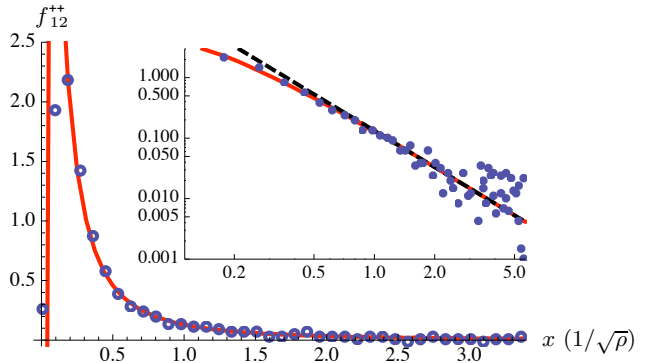}
\caption{Cross-sectional plot of the data points versus theoretical
curves of the pair correlation function (Fig.~\ref{Fig:DensityPlot})
along $\hvect{x}$ axis. After a short distance away from the core, the
function has a power law decay of $1/x^2$ as shown with the dashed
line in the log-log plot in the inset.}
\label{Fig:XSection}
\end{center}
\end{figure}

To verify that Eq.~(\ref{E:fsoln}) is applicable in glide-controlled
systems, we consider an ensemble of 1500 relaxed configurations of 64
plus and 64 minus dislocations randomly placed on a 1~$\mu$m$^2$ square and restricted to
move along their glide directions. The simulations
were performed with periodic boundary conditions in the absence of
thermal noise. The glide constraint helps prevent dislocation
annihilation, and thus, to fix the total number of dislocations and to
maintain the finite effective temperature.
As an example, Fig.~\ref{Fig:DensityPlot} shows (b) the density plot
of the theoretical correlation function $\ff{++}{12}$ between plus dislocations on
$60^\circ$ and $120^\circ$ slip systems against (a) the simulation result.
The erroneous oscillations in Fig.~\ref{Fig:DensityPlot}(b) along
$0^\circ$ and $90^\circ$ lines are caused by the numerical inverse
Fourier transform operation of Eq.~(\ref{E:fsoln}). (The general
closed form solution of a double-slip pair correlation function does
not exist for an arbitrary pair of slip orientation angles.)  Overall,
the theory gives accurate angular predictions except
along the two slip directions where it underpredicts the same-sign
anti-correlation due to the suppression of climb. The plot of the correlation function along the $\hvect{x}$ axis is shown in Fig.~\ref{Fig:XSection}. Very close to the origin, the function diverges logarithmically as does the unscreened potential. About one dislocation spacing from the core, the correlation function decays as $1/x^2$.

The real-space solution to Eq.~(\ref{E:fsoln}) is possible if we assume that \emph{the angle between each adjacent pair
of slip planes is constant}. For any $N \in \mathbf{Z}^+$ and $N>1$,
\begin{equation}
  \sum_{n=1}^N \sin^2\!\left(\phi_k-\frac{n\pi}{N}\right) = \frac{N}{2}\,,
\end{equation}
regardless of $\phi_k$. With the above identity, the denominator of
$\fF_{ij}$ becomes angular independent and can be integrated
directly. The final result, with 
\[
k_0 \equiv \sqrt{2\pi N\Gamma\rho}\,, 
\]
reads
\begin{multline}\label{E:ffinal}
  \ftwo_{ij}(r,\phi)\! = \frac{-2}{\lambda_j}\bigg\{\frac{\cos(2\phi-\theta_i-\theta_j)}{(k_0 r)^2}
- \frac{\cos(\theta_i-\theta_j)}{k_0 r} K_1(k_0 r) \\ -\sin(\phi-\theta_i)\sin(\phi-\theta_j)\,K_2(k_0 r)\bigg\}.
\end{multline}

At large distances, the first term dominates and the pair correlation decays like $1/r^2$ (except along the directions where the argument of the cosine is $\pi/2$, $3\pi/2$, etc.). Compared to the single slip case (Eq.~\ref{E:fsingleslip}) where the pair correlation diminishes exponentially (except along the dislocation wall direction), the presence of extra slip(s) suppresses the Debye screening. 
It should be noted that $-\ftwo_{ij}(\r)$ can be thought of as the effective interaction potential due to screening. More precisely, $\vect{F}^{\rm PK} \sim \vectsym{\nabla} \ftwo_{ij}$ is the Peach--Koehler force felt by a positive dislocation on slip system $i$ due to the induced screening of dislocations on slip system $j$. It has been shown\cite{GromGyorKocs06} that, for single-slip system, the attractive parabolic potential in the glide direction (taken to be along $\hvect{x}$) falls off with a prefactor of $1/|y|^{5/2}$ along the wall direction. Series expansion of $\phi$ in Eq.~(\ref{E:ffinal}) about $\theta_i$ and $\theta_j$ reveals that, for multiple-slip system, the prefactor of the parabolic potential about the glide directions decays as $1/r^2$---slightly more slowly than the single-slip case.
This could explain the necessity to include more than one slip system to see the formation of cell walls and grain boundaries in two-dimensional discrete dislocation simulations prohibiting climb motion.\cite{BenzBrecNeed04,BenzBrecNeed05,FourSala96,GomeDeviKubi06,GromBako00,GromPawl93PMA,GromPawl93MSEA,GullHart93} The analysis also confirms the ``directional long-range order'' of two-dimensional crystals as rigorously proven by Mermin.\cite{Merm68}

\section{Derivation of a multiple-slip evolution law}\label{S:EvolutionLaw}
To arrive at a set of transport equations for an ensemble of multiple-slip dislocation systems, we extend the treatments of Groma et al. in Ref.~\onlinecite{Grom97}, \onlinecite{GromBalo99}, and \onlinecite{ZaisMiguGrom01}. The evolution equations for the uncorrelated single-dislocation densities on slip system $i$ read:
\begin{subequations}\label{E:drhodt}
\begin{align}
\begin{split} \label{E:drhodt1}
\partial_t\rho^\texttt{+}_i&(\ra,t) = -( \vect{b}_i\cdot\vectsym{\nabla})\! \Bigg[ \!\!+\rho^\texttt{+}_i(\ra,t) \tau_i^{\text{ext}} \\
&+ \sum_j \int d^2\rb\,
\left( \rhoss{++}(\ra,\rb,t) - \rhoss{+-}(\ra,\rb,t) \right) \tau_{ij}^{\text{ind}} \Bigg],
\end{split}\\
\begin{split} \label{E:drhodt2}
\partial_t\rho^\texttt{-}_i&(\ra,t) = - ( \vect{b}_i\cdot\vectsym{\nabla})\! \Bigg[ \!\!-\rho^\texttt{-}_i(\ra,t) \tau_i^{\text{ext}} \\
&+ \sum_j \int d^2\rb\,
\left( \rhoss{--}(\ra,\rb,t) - \rhoss{-+}(\ra,\rb,t) \right) \tau_{ij}^{\text{ind}} \Bigg],
\end{split}
\end{align}
\end{subequations}
where the dislocation mobility has been absorbed into the rescaling of time $t$.
With the assumption that all dislocations have the same magnitude $b$, the
Burgers vector can be written as $\vect{b}_i=b \hvect{s}_i$
($\hvect{s}_i$ and $\hvect{m}_i$ respectively are the slip direction and slip plane
normal direction of slip system $i$).
$\tau_{ij}^{\text{ind}}(\ra-\rb)$ is the resolved shear stress exerted on a dislocation
at $\ra$ on slip $i$ by a dislocation at $\rb$ on slip $j$, and can be written as
\begin{equation}\label{E:tau}
  \tau_{ij}^{\text{ind}}(\r) = \hvect{s}_i\cdot\tensor{\sigma}_j\cdot\hvect{m}_i = G\,b\,(\hvect{s}_i\cdot\vectsym{\nabla})(\hvect{m}_i\cdot\vectsym{\nabla})(\hvect{m}_j\cdot\vectsym{\nabla}) \! \left[ r^2\ln r \right]\!.
\end{equation}
Here, $G \equiv \mu /(2\pi(1-\nu)) = E/(4\pi(1-\nu^2))$, where $E$, $\mu$, $\nu$ are the Young's modulus, shear modulus, and Poisson ratio respectively.

Addition and subtraction of Eqs. (\ref{E:drhodt1}) and (\ref{E:drhodt2}) give the evolution equations for the total dislocation density $\rho_i \equiv \rho_i^\texttt{+} + \rho_i^\texttt{-}$ and the GND density $\kappa_i \equiv \rho_i^\texttt{+} - \rho_i^\texttt{-}$:
\begin{subequations}\label{E:rho_kappa_evol}
\begin{align}
\begin{split}
\partial_t\rho_i &= - (\vect{b}_i\cdot\vectsym{\nabla})\! \Bigg[
\kappa_i\tau_i^{\text{ext}} \\
&+ \sum_j \int d^2\rb \underbrace{ \left(\rhoss{++} + \rhoss{--} - \rhoss{+-} - \rhoss{-+} \right)}_{\equiv \kappa_{ij}^{(2)}(\ra,\rb,t)} \tau_{ij}^{\text{ind}} \Bigg] \label{E:rho_evol}
\end{split}\\
\begin{split}
\partial_t\kappa_i &= - (\vect{b}_i\cdot\vectsym{\nabla})\! \Bigg[
\rho_i\tau_i^{\text{ext}} \\
&+ \sum_j \int d^2\rb \underbrace{ \left(\rhoss{++} - \rhoss{--} - \rhoss{+-} + \rhoss{-+} \right)}_{\equiv \rho_{ij}^{(2)}(\ra,\rb,t)} \tau_{ij}^{\text{ind}} \Bigg] \label{E:kappa_evol}
\end{split}
\end{align}
\end{subequations}
In accordance with (\ref{E:g}), 
the dislocation--dislocation density can be written as
\begin{equation}\label{E:rho_rho}
\begin{split}
\rho^{ss'}_{ij} &= \rho_i^s(\ra)\rho_j^{s'}(\rb)g_{ij}^{ss'}(\ra-\rb) \\
&= \rho_i^s(\ra)\rho_j^{s'}(\rb)(1 + d_{ij}^{ss'}(\ra-\rb)) \,,
\end{split}
\end{equation}
where $s,s'\in \{+,-\}$ and, according to (\ref{E:g2}), $d^{ss'}_{ij} = \Gamma f^{ss'}_{\,ij}$.
In terms of the single and pair correlation functions, the total dislocation density and GND are
\begin{subequations}\label{E:rho2_kappa2}
\begin{align}
\begin{split}
\rho_{ij}^{(2)} =\, &\rho_i(\ra)\rho_j(\rb) + \frac{1}{2} \Big\{-\rho_i(\ra)\rho_j(\rb) d^a_{ij} \\
&+ \rho_i(\ra)\kappa_j(\rb)[d^p_{ij} + d^s_{ij}] \\ &+\kappa_i(\ra)\rho_j(\rb)[d^p_{ij}-d^s_{ij}] + \kappa_i(\ra)\kappa_j(\rb)d^a_{ij}\Big\},
\end{split}\\
\begin{split}
\kappa_{ij}^{(2)} =\, &\kappa_i(\ra)\kappa_j(\rb) + \frac{1}{2} \Big\{ \rho_i(\ra)\rho_j(\rb) [ d^p_{ij} - d^s_{ij}] \\
&+ \rho_i(\ra)\kappa_j(\rb)d^a_{ij} -\kappa_i(\ra)\rho_j(\rb)d^a_{ij} \\
&\qquad+ \kappa_i(\ra)\kappa_j(\rb)[d^p_{ij}+d^s_{ij}]\Big\},
\end{split}
\end{align}
\end{subequations}
where $d^p_{ij} = \dss{++}$, $d^s_{ij} = (1/2)(\dss{+-} + \dss{-+})$, and $d^a_{ij} = (1/2)(\dss{+-} - \dss{-+})$.
After substitution of Eqs.~(\ref{E:rho_rho})--(\ref{E:rho2_kappa2}), Eq.~(\ref{E:rho_kappa_evol}) becomes
\begin{subequations}\label{E:rho_kappa_evol2}
\begin{align}
\partial_t\rho_i &= - (\vect{b}_i\cdot\vectsym{\nabla})\! \left[ \kappa_i(\tau_i^{\text{ext}} + \tau_i^{\text{sc}} - \tau_i^\text{f} - \tau_i^\text{b}) + \rho_i\tau_i^\text{a} \right], \label{E:rho_evol2} \\
\partial_t\kappa_i &= - (\vect{b}_i\cdot\vectsym{\nabla})\! \left[ \rho_i(\tau_i^{\text{ext}} + \tau_i^{\text{sc}} - \tau_i^\text{f} - \tau_i^\text{b}) + \kappa_i\tau_i^\text{a} \right], \label{E:kappa_evol2}
\end{align}
\end{subequations}
in which 
\begin{subequations}
\begin{align}
\tau_i^\text{sc} &= \sum_j \int \kappa_j(\rb)\tau_{ij}^{\text{ind}}(\ra-\rb)\, d^2\rb, \\
\tau_i^\text{b} &= -\frac{1}{2} \sum_j \int \kappa_j(\rb)d^t_{ij} \tau_{ij}^{\text{ind}}(\ra-\rb)\, d^2\rb, \label{E:tau_b1}\\
\tau_i^\text{f} &= \frac{1}{2} \sum_j \int \rho_j(\rb)d^a_{ij} \tau_{ij}^{\text{ind}}(\ra-\rb)\, d^2\rb, \label{E:tau_f} \\
\begin{split}
\tau_i^\text{a} &= \frac{1}{2}\sum_j \int \rho_j(\rb)[d^p_{ij} - d^s_{ij}]\tau_{ij}^{\text{ind}}(\ra-\rb) \\
&\qquad\qquad\qquad+ \kappa_j(\rb)d^a_{ij} \tau_{ij}^{\text{ind}}(\ra-\rb) \, d^2\rb. \label{E:tau_a}
\end{split}
\end{align}
\end{subequations}
The term $d^t_{ij} \equiv (1/4)(\dss{++} + \dss{--} + \dss{+-} + \dss{-+})$ in (\ref{E:tau_b1}) involves averaging over pairs of correlation functions.

Terms involving $\tau^\text{a}_{i}$ in Eq.~(\ref{E:rho_kappa_evol2}) can be cast away by going into a ``co-moving'' frame of $\rho_i$ and $\kappa_i$ respectively. Although $\dss{++} = \dss{--} = -\dss{+-} = -\dss{-+}$ and hence $d^t_{ij}$ should vanish by definition, this is hardly the case when, e.g., the system is strained through external loading. Only one of these correlation functions dominates locally, resulting in a nonzero $d^t_{ij}$. Similarly the contribution from flow stress, $\tau^\text{f}_{i}$, is greatest in regions with equal population of plus and minus dislocations; in most regions, its effect is negligible. We shall therefore focus only on the contribution from back stress $\tau_i^\text{b}$. The validity of this assumption is supported by the success of the recent single-slip theory.\cite{YefiGromGies04,YefiGromGies04b}

Although $d^t_{ij}(\r)$ is long-range, the magnitude of the back stress $\tau_i^\text{b}$ is still considerably smaller than that of the self-consistent internal stress $\tau_i^\text{sc}$ when $r$ is large compared with mean dislocation spacing. We are therefore interested in the contribution of $d^t_{ij}(\r)$ to the stress only at short distances where its effect is much more pronounced. Consider a dislocation at $\ra$, we can Taylor expand $\kappa_j(\rb)$ about this point, $\kappa_j(\rb) \simeq \kappa_j(\ra) + (\rb-\ra)\cdot\vectsym{\nabla} \kappa_j\Big|_{\ra} +$~terms of higher orders. Because $d^t_{ij}(\r)$ is symmetric while $\tau_{ij}^{\text{ind}}(\r)$ is anti-symmetric under $\r\mapsto -\r$, the first term in the expansion vanishes. We then make a change of variable to the scaled coordinate $\sqrt{\rho}\,\r \mapsto \vec{x}$, where $\rho$ represents the mean total dislocations of the system. To second order this yields
\begin{equation}\label{E:tau_b}
\tau_i^\text{b}(\ra) = \sum_{j=1}^N \frac{\vectsym{\nabla}\kappa_j}{\rho} \cdot \int \vect{x}\, d^t_{ij}(\vect{x})\tau_{ij}^{\text{ind}}(\vect{x}) d^2\vect{x}\,.
\end{equation}

Using the Fourier transform expression of $d^t_{ij}$, the integral in Eq.~(\ref{E:tau_b}) can be evaluated directly using Parseval's theorem:
\begin{equation}\label{E:Iint}
\vect{I}_{ij} \equiv \int \vect{x}\, d^t_{ij}(\vect{x},\thab)\tau_{ij}^{\text{ind}}(\vect{x}) d^2\vect{x} = \int \dtF(\vect{k})\,\mathcal{F}\!\left[\vect{x}\,\tau_{ij}^\text{ind}\right]\![\vect{k}]\,d^2\vect{k}
\end{equation}
The Fourier transform of $\vect{x}\,\tau_{ij}^\text{ind}$ can be computed directly from (\ref{E:tau}):
\begin{equation}\label{E:tauF}
\begin{split}
  \mathcal{F}\!\left[\vect{x}\,\tau_{ij}^\text{ind}\right]\![\vect{k}] &= -4\pi G\,b\, \Dk\!\left[\frac{ (\hvect{s}_i\cdot\vect{k})(\hvect{m}_i\cdot\vect{k})(\hvect{m}_j\cdot\vect{k})}{k^4}\right] \\
&= -G\,b\Dk\!\left[(\hvect{s}_i\cdot\vect{k})\psiF_{ij} \right]
\end{split}
\end{equation}
Owing to the connection
$d^t_{ij}(\vect{x}) = \Gamma\,\ftwo_{ij}(\vect{x})$, Eq.~(\ref{E:Iint}) becomes, from (\ref{E:fsoln}) and (\ref{E:tauF}),
\begin{equation}\label{E:Imidstep}
  \vect{I}_{ij} = \frac{\Gamma^2 G\,b}{\lambda_j} \int \frac{\psiF_{ij} \Dk\!\left[(\hvect{s}_i\cdot\vect{k})\psiF_{ij} \right]}{1-\sum_n \psiF_{nn}}\, d^2\vect{k}.
\end{equation}
\begin{widetext}
The vector $\vect{I}_{ij}$ is most conveniently expressed in the
coordinate system of slip $j$. Substitution of Eq.~(\ref{E:psi}) into Eq.~(\ref{E:Imidstep}), while projecting $\hvect{s}_i$ and $\hvect{m}_i$ onto $(\hvect{s}_j,\hvect{m}_j)$, gives
\begin{multline}
  \vect{I}_{ij} = (4\pi)^2\frac{\Gamma^2 G\,b}{\lambda_j} \Bigg\{\hvect{s}_j \int_0^{2\pi} \int_\epsilon^\infty \frac{-1}{k} \frac{\sin^2(\phi_k)\sin(\phi_k+\theta_{ij})\sin(3\phi_k+2\theta_{ij})}{k^2+4\pi\sum_n \sin^2(\phi_k-\theta_n)}\, dk\,d\phi_k\\
+\hvect{m}_j \int_0^{2\pi}\int_\epsilon^\infty \frac{1}{2k}\frac{\sin(\phi_k)\sin(\phi_k+\theta_{ij})\sin(4\phi_k+2\theta_{ij})}{k^2+4\pi\sum_n \sin^2(\phi_k-\theta_n)}\, dk\,d\phi_k \Bigg\},
\end{multline}
where $\theta_{ij} = (j-i)\pi/N$ is the angle between slip planes $i$ and $j$.
We impose a cut-off $\epsilon$ at small $k$ to prevent the logarithmic divergence due to the long-range nature of the pair correlation functions.
\end{widetext}

Under the assumption of equal interval of successive slip orientation, as in
the previous section, we can carry out the above integrals very straightforwardly, giving
\begin{equation}
  \vec{I}_{ij} = \frac{GD\,b}{\lambda_j}\cos(\theta_{ij})\hvect{s}_j
\end{equation}
where $D = 2\pi^2\Gamma^2|\ln\epsilon|/N$ serves as a fitting parameter. The factor $\lambda_j$ nicely combines with $\rho$ in the denominator of Eq.~(\ref{E:taub}) to make $\rho_j = \lambda_j \rho$. For physical reasons, we are going to replace $\rho_j$ with its local density $\rho_j(\r)$. In the previous sections, we calculated the pair correlation functions of an ensemble of \emph{spatially constant} single-dislocation densities in thermal equilibrium. When the distributions of single-dislocation densities are non-uniform in space as is the case for systems out of equilibrium, the back stress response should depend on how much the densities vary locally.

The final result is amazingly simple:
\begin{equation}\label{E:taub}
  \tau_i^\text{b}(\r) = GD \sum_{j=1}^N \cos(\theta_{ij}) \frac{(\vect{b}_j\cdot\vectsym{\nabla})\kappa_j(\r)}{\rho_j(\r)}
\end{equation}
The above form for the back stress converges nicely to the single-slip theory of Groma~et al.\cite{GromCsikZais03,YefiGromGies04,YefiGromGies04b,YefiGies05,YefiGies05b}
The $\cos(\theta_{ij})$ coupling between slip systems should come as no surprise. The angular dependence of the back stress must emerge from the symmetry of the potential. The angular average of $\psi_{ij}$ in Eq.~(\ref{E:IntEnergy}) selects out $\cos(\theta_{ij})$ as the only possibility.
It is interesting to note also that the same coupling
also appears in the strain gradient theory for continuum crystal plasticity by Gurtin.\cite{Gurt00,Gurt02,Gurt03}

\section{Comparison with the earlier multislip plasticity theory}\label{S:Comparison}
Recently, Yefimov et al.\cite{YefiGies05,YefiGies05b} have proposed an extension of their single-slip continuum plasticity theory\cite{YefiGromGies04,YefiGromGies04b} to incorporate systems with more than one slip. In their theory, each slip system $j$ contributes some amount of back stress, given in our notation by
\begin{equation}
  \tau_j^\text{b}(\r) = GD \frac{(\vect{b}_j\cdot\vectsym{\nabla})\kappa_j(\r)}{\rho_j(\r)}
\end{equation}
to the total back stress of slip system $i$ according to
\begin{equation}
  \tau_i^\text{tot} = \sum_{j=1}^N S_{ij}\tau_j^\text{b}
\end{equation}
with slip-orientation dependent weight factor $S_{ij}$ acting as a
projection matrix. For symmetry reason, three variations were postulated:\cite{YefiGies05,YefiGies05b}
\begin{subequations}
\begin{align}
S^1_{ij} &= (\hvect{m}_i\cdot\hvect{m}_j)(\hvect{s}_i\cdot\hvect{s}_j) = \cos^2(\theta_{ij}) \\
S^2_{ij} &= \hvect{m}_i\cdot(\hvect{s}_j\otimes\hvect{m}_j + \hvect{m}_j\otimes\hvect{s}_j)\cdot\hvect{s}_i = \cos(2\theta_{ij}) \label{E:whattheychose} \\
S^3_{ij} &= \hvect{s}_i\cdot\hvect{s}_j = \cos(\theta_{ij}) \label{E:sameasours}
\end{align}
\end{subequations}
Note that the third possibility (\ref{E:sameasours}) is consistent with
the expression for the back stress we have derived in (\ref{E:taub}).

To select among these choices, Yefimov et al. successively used all
three laws to numerically analyze the problem of simple shearing of a crystalline strip containing two slip systems with impenetrable walls.\cite{YefiGies05} The results of each case were compared against that from the discrete dislocation simulations of Shu et al.\cite{ShuFlecGiesNeed01} The best match was achieved with Eq.~(\ref{E:whattheychose}). Other choices underpredicted the amount of plastic strain. The chosen interaction law was then tested against the problem of bending of a single crystal strip with satisfactory agreement with discrete dislocation results of Cleveringa et al.\cite{ClevGiesNeed99}

We believe that the success of their continuum theory in the shearing
problem despite the incorrect choice of interaction law is due to a
different reason. The amount of plastic strain is controlled by (i)
the fitting parameter $D$ and (ii) the number density of nucleation sites in the film. By adjusting these values, different interaction laws could be altered to obtain the desired fit. In their analysis, Yefimov et al. used the value of $D$ from their previous single-slip theory\cite{YefiGromGies04} without any readjustment. There is no a priori reason why this value should stay unaltered. The density of nucleation sources in their continuum theory were chosen to match that in the discrete dislocation simulations. The discrepancy could also arise from different ways in which the discrete dislocation theory and the continuum theory handle dislocation nucleation.

In a later publication, Yefimov et al. applied their formalism to the
problem of stress relaxation in single-crystal thin films on
substrates subjected to thermal loading.\cite{YefiGies05b} Due to the
difference in thermal expansion coefficients between film and
substrate, high tensile stresses can develop in the films as the
temperature decreases. Contrary to the discrete dislocation
simulations by Nicola et al.\cite{NicoGiesNeed03,NicoGiesNeed05} which
show increasing stress built up inside a film with decreasing film
thickness, the results from the continuum theory show a size-dependent hardening only during the early stage of cooling. Moreover, the theory gives identical results between some pair of slip orientations (e.g. when the angle between the two slip planes $\theta_{12}$ is either $60^\circ$ or $120^\circ$), whereas the discrete dislocation simulations and our new theory predict otherwise.
Finally, in the previous continuum theory,\cite{YefiGies05} dislocations nucleate when the sum of the external stress $\tau^\text{ext}$, the self-consistent long-range stress $\tau^\text{sc}$, and back stress $\tau^\text{b}$ exceed a certain value. From our analysis, we believe that, in a more correct treatment of dislocation nucleation, this back stress should be supplemented by flow stress $\tau^\text{f}$ (Eq.~(\ref{E:tau_f})) which is dominant in a nucleation region where plus and minus dislocations are equally populated.

Applications of the current theory to the shearing problem and the thin film problem which shows the size-dependent hardening will appear shortly following this publication.

\section{Discussion and conclusions}
We have described $n^\text{th}$-order dislocation densities and dislocation pair
correlation functions in a grand canonical ensemble and obtained the
relationships between different orders of the correlation functions in the
form of a hierarchy of integral equations.
Using the Bogolyubov ansatz instead of the more customary Kirkwood
approximation, we have closed the chain of the equations at second order and solved
for approximate expressions of the pair correlation functions---valid at all
distances---for systems with one slip and multiple active slip
systems. These solutions are invariant under simultaneous
transformations $\r \mapsto \r/\sqrt{\rho}$ and $\rho \mapsto
\rho^2$. The transformations suggest that any emergent dislocation
pattern should exhibit a length scale given by $1/\sqrt{\rho}$ as
pointed out by Holt,\cite{Holt70} and in agreement with the ``law of
similitude.''\cite{RajPhar86} For a complete analysis of scaling
relations the reader is referred to Ref.~\onlinecite{ZaisMiguGrom01}.

Recently Groma~et~al.~have developed a mean-field variational approach to study
the screening of dislocations,\cite{GromGyorKocs06} similar in spirit to the
Debye--H{\"u}ckel theory in the study of classical
plasmas.\cite{DebyHuck23,LandLifs60,LandLifs69} This method is based on
approximating the system's total density matrix as a product of single-particle density matrices $\rho_i$ with the free energy given by $F = \avg{\mathcal{H}} + T \sum_i \text{Tr} \rho_i \ln\rho_i$. Although this technique provides a complimentary approach and results in the same pair correlation expressions for a single-slip system (after some interpretation), its generalization to multiple-slip system is not obvious. In particular, one would have to supply additional cross couplings between different slips by hand. These couplings should automatically emerge from a complete theory.

In Sec.~\ref{S:EvolutionLaw}, we have formulated transport equations for the total
dislocation and GND densities for general multiple slip. Interactions among
dislocation pairs produce an additional (relatively) short-ranged ``back
stress'' contribution to the long-range internal stress of individual
dislocations. Most of the complexities of the correlation functions were
integrated away, leaving only the $\cos(\theta_{ij})$ coupling between slip
systems $i$ and $j$, see Eq.~(\ref{E:taub}). This dependence was also
proposed by Gurtin in his strain gradient plasticity
theory.\cite{Gurt02,Gurt03} but was abandoned by
Yefimov~et~al.\cite{YefiGies05,YefiGies05b} We have argued in
Sec.~\ref{S:Comparison} that this refusal was based on an unfair
comparison with discrete dislocation simulations for the way in which
dislocation nuncleation was treated. 

There is an important issue regarding the use of dislocation correlations $\ftwo_{ij}$ for $d^t_{ij}$ in Sec.~\ref{S:MultiSlip}.
The formalism developed in Sec.~\ref{S:BBGYK} assumes that dislocations relax
along the directions dictated by Peach--Koehler forces. This implies
dislocation glide, as included in the transport equations developed in
Sec.~\ref{S:EvolutionLaw}, but also climb which is not considered a mechanism
of plastic flow here. Mathematically speaking, Eq.~(\ref{E:BBGYKrho}) is
\emph{not} the stationary state of Eq.~(\ref{E:drhodt}). Early
attempts in numerically describing dislocation correlations in glide-only,
multiple-slip systems failed to
produce noticable patterns due to the need for large number of
dislocations; the role of climb (or cross slip) was suggested to help overcome this
difficulty.\cite{BakoGromGyorZima06,BakoGromGyorZima07} The original
motivation for our approach was to find the orientation dependence of the back stress in the most straightforward way. Extracting the angular dependence from a climb-assisted relaxed state gave us a quick input to use in the glide-only multiple-slip theory. The validity of the continuum theory will always be vindicated by comparisons against discrete dislocation results.

Finally, we believe that our multiple-slip formulation provides a framework to
address a long standing challenge in explaining dislocation patterning. For
single-slip systems, short-range correlations occur between two dislocations
except along directions normal to their glide plane (taken to be along
$\hvect{y}$). It has been shown that for a small deviation away from this
``dislocation wall'' direction, an attractive parabolic potential produced by the
correlated dislocations decays as $|y|^{-5/2}$, compared
with $|y|^{-2}$ in the unscreened case.\cite{GromGyorKocs06} We have found in Sec.~\ref{S:MultiSlip}, however, that when one or more extra slips are introduced, the effect of Debye-like screening
diminishes. In this case, the attractive potential in fact decays like
$r^{-2}$ as if it were unscreened. This could explain the necessity to
introduce extra slips to see the formation of walls in discrete dislocation
simulations,\cite{BenzBrecNeed04,BenzBrecNeed05,FourSala96,GomeDeviKubi06,GromBako00,GromPawl93PMA,GromPawl93MSEA,GullHart93}
\emph{unless} further aided by climb
motions.\cite{BartCarl97,BakoGromGyorZima06} The latter suggests the existence
of a critical exponent of the attractive potential below which structure
formation cannot occur as is the case in single-slip systems restricted to
glide. A more detailed investigation of this is left for future work.

\begin{acknowledgments}
The authors are grateful to Professor Istv\'an Groma for his insightful input
and valuable suggestions. We also would like to thank P\'eter Dus\'an Isp\'anovity for providing us with the discrete dislocation dynamics data used in Sec.~\ref{S:MultiSlip}. Funding from the European Commissions
Human Potential Programme \textsc{SizeDepEn} under contract number MRTN-CT-2003-504634 is acknowledged.
\end{acknowledgments}

\bibliography{references.bib}

\end{document}